\documentclass[prl,twocolumn,superscriptaddress,noshowpacs,nobalancelastpage]{revtex4}

\usepackage[pdftex]{graphicx}
\usepackage{graphicx}
\usepackage{dcolumn}
\usepackage{bm}
\usepackage{siunitx}

\DeclareSIUnit{\belmilliwatt}{Bm}
\DeclareSIUnit{\dBm}{\deci\belmilliwatt}
\newcommand{\ket}[1]{\left| #1 \right>} 
\begin{document}
\bibliographystyle{apsrev4-1}

\preprint{APS}

\title{Gate-Based High Fidelity Spin Readout in a CMOS Device}

\author{Matias Urdampilleta	}
\email{matias.urdampilleta@neel.cnrs.fr}
\affiliation{Institut N\'eel, CNRS/Universit\'e Grenoble Alpes, 38402 Grenoble, France}

\author{David J. Niegemann	}
\affiliation{Institut N\'eel, CNRS/Universit\'e Grenoble Alpes, 38402 Grenoble, France}

\author{Emmanuel Chanrion}
\affiliation{Institut N\'eel, CNRS/Universit\'e Grenoble Alpes, 38402 Grenoble, France}

\author{Baptiste Jadot}
\affiliation{Institut N\'eel, CNRS/Universit\'e Grenoble Alpes, 38402 Grenoble, France}

\author{Cameron Spence}
\affiliation{Institut N\'eel, CNRS/Universit\'e Grenoble Alpes, 38402 Grenoble, France}

\author{Pierre-Andr\'e Mortemousque}
\affiliation{Institut N\'eel, CNRS/Universit\'e Grenoble Alpes, 38402 Grenoble, France}

\author{Christopher B{\"a}uerle}
\affiliation{Institut N\'eel, CNRS/Universit\'e Grenoble Alpes, 38402 Grenoble, France}

\author{Louis Hutin}
\affiliation{CEA, LETI, Minatec Campus, F-38054 Grenoble, France}

\author{Benoit Bertrand}
\affiliation{CEA, LETI, Minatec Campus, F-38054 Grenoble, France}

\author{Sylvain Barraud}
\affiliation{CEA, LETI, Minatec Campus, F-38054 Grenoble, France}

\author{Romain Maurand}
\affiliation{CEA, INAC-PHELIQS, F-38054 Grenoble, France}

\author{Marc Sanquer}
\affiliation{CEA, INAC-PHELIQS, F-38054 Grenoble, France}

\author{Xavier Jehl}
\affiliation{CEA, INAC-PHELIQS, F-38054 Grenoble, France}

\author{Silvano De Franceschi }
\affiliation{CEA, INAC-PHELIQS, F-38054 Grenoble, France}

\author{Maud Vinet}
\affiliation{CEA, LETI, Minatec Campus, F-38054 Grenoble, France}

\author{Tristan Meunier}
\email{tristan.meunier@neel.cnrs.fr}
\affiliation{Institut N\'eel, CNRS/Universit\'e Grenoble Alpes, 38402 Grenoble, France}

\date{\today}
\begin{abstract}
The engineering of electron spin qubits in a compact unit cell embedding all quantum functionalities is mandatory for large scale integration. In particular, the development of a high-fidelity and scalable spin readout method remains an open challenge.
Here we demonstrate high-fidelity and robust spin readout based on gate reflectometry in a CMOS device comprising one qubit dot and one ancillary dot coupled to an electron reservoir to perform readout. 
This scalable method allows us to read out a spin with a fidelity above 99\% for 1~\si{\milli\second} integration time. 
To achieve such fidelity, we exploit a latched spin blockade mechanism that requires electron exchange between the ancillary dot and the reservoir. 
We show that the demonstrated high read-out fidelity is fully preserved up to 0.5~\si{\kelvin}. This results holds particular relevance for the future co-integration  of spin qubits and classical control electronics.

\end{abstract}

\maketitle

\section{\label{sec:level1}Introduction}

Most of the proposed architectures for large-scale quantum information processing rely on the so-called Surface Code proposal which consists in a two-dimensional (2D) arrangement of qubits \cite{Fowler2011}. 
Electron spins in silicon nanodevices offer the advantages to be compatible with modern microelectronic fabrication \cite{Maurand2016}, and to present high fidelity gate operations with a quiet environment \cite{Veldhorst2015, Zajac2018, Watson2018, yoneda2018quantum}. 
This two aspects open a promising road to scale up quantum architecture on a chip where electron spins are stored in arrays of quantum dots \cite{Vandersypen2017,Veldhorst2017,usineagaz}. 
Even though control of small 2D array have been achieved \cite{Thalineau2012, Flentje2017, Mukhopadhyay2018, Mortemousque2018}, the problem of scalability imposes severe constrains on the gate layout \cite{li2018crossbar, Veldhorst2017}, the positioning of the electron reservoirs and the charge readout strategy. 
Modern complementary-metal-oxide-semiconductor (CMOS) technology offers the possibility to relax this constrains by fabricating multilayer devices where local reservoirs and detectors can be implemented \cite{usineagaz}. 
However, this strategy requires the development of new functionalities, especially for the readout procedure. 
The single shot detection of electron spins in semiconductor quantum dots is based on a spin-to-charge conversion achieved traditionally through two different methods. The first method is the so-called energy selective tunneling which consists in the spin dependent tunneling of an electron between the qubit and a reservoir \cite{Elzerman2004}. The second method relies on the Pauli spin blockade (PSB) effect. It requires the presence of a second quantum dot, the so-called readout ancillary dot, which also contains a single electron. The PSB prevents the two electrons from tunneling on the same dot if they have parallel spin orientation \cite{ono2002current, PhysRevLett.103.160503, Neeley2009}. To measure the spin of a single electron in the qubit dot, the electron in the readout ancillary dot has to be initialized in the $\ket{\downarrow}$ ground state. However, these two methods require an external charge detector to convert the spin information into an electrical signal. Having a local reservoir and a charge sensor for every qubit precludes those methods from being integrated in large scale quantum architectures. 
In the present letter, we propose and demonstrate a new readout method which reduces considerably the number of nanoscale components needed for readout and presents fidelities above $99$\% compatible with large-scale quantum information processing. 
It consists in combining a radio-frequency (RF) gate reflectometry technique \cite{petersson2010charge} with an electron latching mechanism \cite{fogarty2017integrated, Harvey}. 
In this configuration, the charge detector, made of a dot tunnel-coupled to one reservoir and connected to a RF gate reflectometry set-up, goes beyond its standard role as it participates in the spin-to-charge conversion through a spin-blockade mechanism. 
Moreover, we have achieved this demonstration using a $300$~\si{\milli\metre} compatible CMOS device to ensure its large-scale integration. 
We have studied the relaxation process in such a configuration and demonstrate its robustness with respect to temperature with high fidelities up to $500$~\si{\milli\kelvin}.
\begin{figure*}[t]%
\includegraphics[width=15cm]{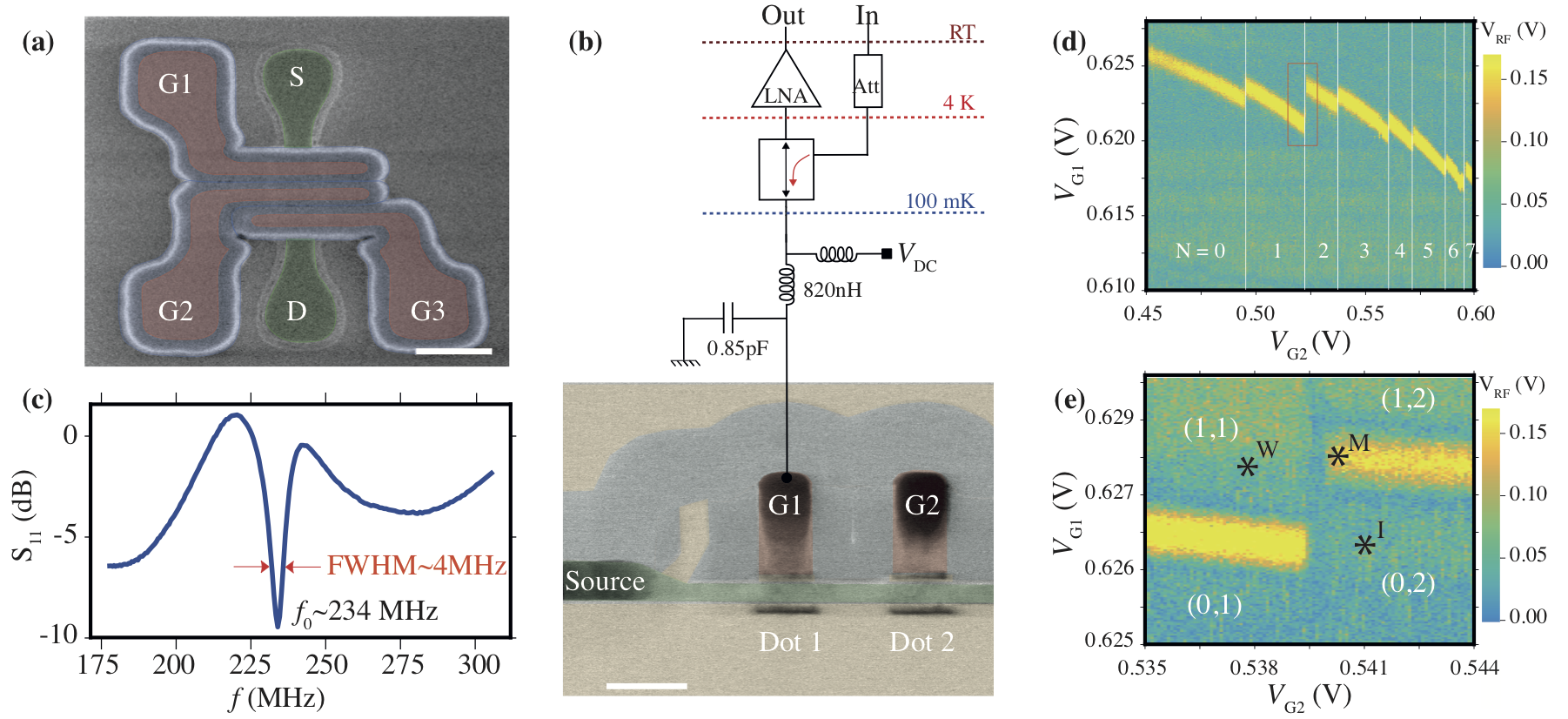}%
\caption{ \textbf{A CMOS device probed by gate-based RF reflectometry.} 
(a) SEM micrograph of the CMOS device. The silicon wire (green) lays on the buried oxide layer and is covered with the three top gates (brown). The scale bar corresponds to $200$~\si{\nano\metre}. 
(b) The upper panel shows the reflectometry setup which comprises a tank circuit composed of an inductance and the parasitic capacitance to ground. The injected signal once reflected is further amplified using a cryogenic amplifier and demodulated at room temperature. 
The lower panel shows a TEM micrograph of a slice of the device. The two dots are located underneath gate 1 and 2. For simplicity the gate 3 is not represented as it is not used in the present experiment. The scale bar corresponds to $40$~\si{\nano\metre}. 
(c) Reflected signal from the tank circuit. The resonator presents at low temperature a quality factor of $\approx 50$. 
(d) The amplitude of the reflected signal is plotted as a function of gate 1 and 2 voltages. The dot 1 is strongly coupled to the reservoir, as a consequence, a strong amplitude variation is visible for its charge degeneracy. When the total number of electron in the quantum dot 2 changes, it shifts the chemical potential of the dot 1 as highlighted by the solid white lines, where the number N stands for the number of electrons in the quantum dot 2. 
(e) corresponds to the region framed by the red square in (e), it shows the interdot transition from (0,2) to (1,1).  The points I, W and M are used to respectively  initialize, wait for relaxation and measure the spin.
}
\label{fig:stab}%
\end{figure*}

\section{Charge characterization of the CMOS device}
The triple gate device, sketched in Fig.~\ref{fig:stab}(a) is fabricated from an SOI substrate with standard CMOS technology (details in the Method section). 
Gate 1 is directly connected to a tank circuit to achieve RF reflectometry \cite{Gonzalez-Zalba2015, Urdampilleta2015a} (see Fig.~\ref{fig:stab}(b), (c) and (d)) and probe for the charge configuration of the device. 
We first characterize the charge stability of the device which is operated in a double quantum dot configuration. 
For this purpose, we apply positive voltage on gate 1 and 2 to form the quantum dots at the interface and we apply a negative voltage on gate 3 in order to isolate the quantum dot 2 from the electron reservoir. 
The RF signal is sent on gate 1 and the reflected signal is analyzed. 
The corresponding demodulated signal, see Fig.~\ref{fig:stab}(e), represents the change of amplitude induced by change of capacitance between gate 1 and the channel. 
As a consequence, the broad line is the signature of electrons tunneling between one level of the quantum dot under gate 1 aligned with the Fermi energy. 
As gate 2 is swept, interruptions in the dot 1 charge degeneracy line are observed. 
It corresponds to charging events of the dot 2 which is capacitively coupled to dot 1. 
The dot 1 can then be used as a charge detector as it permits to sense dot 2 occupancy \cite{hile2015radio}. 
Reflectometry sensing cannot be resolved directly on dot 2 as the tunnel coupling with the reservoir is too small compare to the excitation frequency. 
We can conclude that the dot 2 can be emptied at negative voltages on gate 2 and that the few-electron regime has been achieved in our CMOS device. 
From temperature dependence spectroscopy, a lever arm $\alpha$ factor of 0.15 similar for both gates on their respective underneath dots is obtained. 
The charging energy in the few-electron regime is estimated to be $\approx 4$~\si{\milli\electronvolt} ($\approx 1.5$~\si{\milli\electronvolt} in the high number electron regime). 
It is comparable to previously reported measurement in CMOS dot devices \cite{hofheinz2006simple}. 
In the following, we focus on a region depicted in Fig.~\ref{fig:stab}(f) where PSB is used to perform spin readout. 
It corresponds to a $(N,2)$ to $(N+1,1)$ charge transition with $N$ an even number $-$ for simplicity in the following we set $N=0$.

\section{Single-shot spin readout}
We now show how we perform the spin readout without external charge detector. 
First, we consider the different available spin states. 
At the (0,2)-(1,1) transition, due to exchange interaction between the two electrons, the correct spin basis is the so-called singlet (S) and triplet ($\text{T}_0$, $\text{T}_+$, $\text{T}_-$) spin states. 
This basis offers the possibility to convert the spin into charge information through the Pauli spin blockade. 
This conversion is usually performed by preparing a state in the (1,1) charge configuration. 
When the system is pulsed at the interdot transition, the tunneling from (1,1) to (0,2) is only possible if the two electrons form a singlet state. 
In the opposite case, the spin blockade prevents the charge from tunneling in the (0,2) state until the spin relaxes. 
In the present study and in contrast with recent experiments \cite{DzurakReflecto, SimmonsReflecto}, the measurement cannot be performed at the interdot charge transition because of the lack of signal in the single-shot regime. 
\begin{figure}[t]%
\includegraphics[width=\columnwidth]{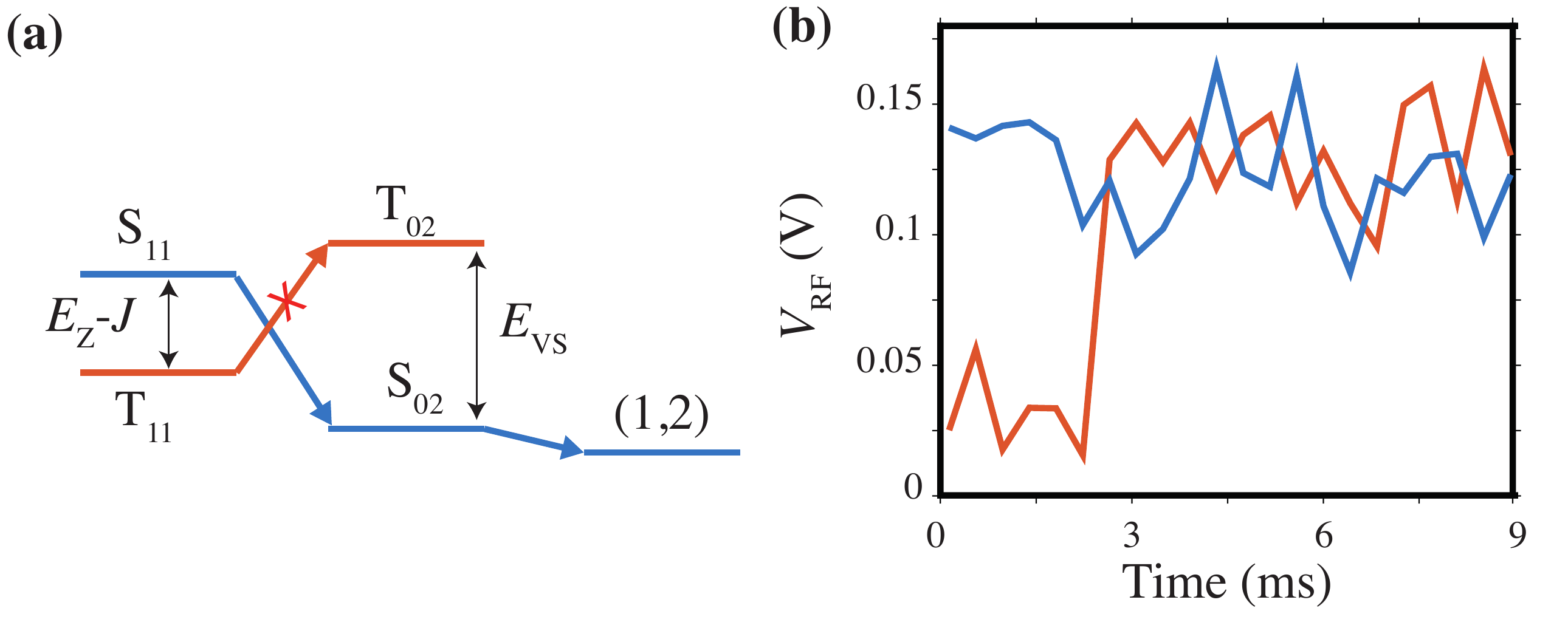}%
\caption{\textbf{Single shot spin readout}. 
(a) Latched spin blockade mechanism. In the (1,1) region, the triplet state $\text{T}_-$ is the ground state when the Zeeman energy ($E_Z$) exceeds the exchange energy $J$. At the measurement point, where the exchange dominates, the singlet can tunnel from the (1,1) to the (0,2) state which in turn can tunnel to the (1,2)  state. On the other hand, triplet state cannot tunnel to the (0,2) state due to spin blockade and the (1,2) state cannot be reached. 
(b) Single shot measurement of the singlet and triplet states. The spin system is initialized in the singlet state and then moved to the W point (see Fig.\ref{sec:level1}(f)). The blue (orange) curve shows typical time trace at the measurement point after short (long) waiting time. The orange curve presents a low signal at short measurement time as the system is stuck in the triplet (1,1) state. Once the triplet state has relaxed to the singlet state the spin blockade is lifted allowing tunneling as shown by the signal at higher level.   
}
\label{fig:latched}%
\end{figure}
To improve the PSB signature, an alternative method has been developed recently \cite{nakajima2017robust, fogarty2017integrated,Harvey}: the so-called latched PSB which involves the tunneling of a third charge, therefore improving the sensitivity of the spin to charge conversion as the total number of charge changes. 
This mechanism relies on the fact that the two quantum dots are not evenly coupled to the electron reservoirs. 
For instance, adding a charge on dot 2 is much slower than for dot 1. 
As a consequence, the charge transition (1,1) to (1,2) is much slower than (0,2) to (1,2). In term of spin state, a triplet (1,1) for instance needs first to turn into a singlet state before tunneling to (0,2) and then to (1,2), see Fig.~\ref{fig:latched}(a). 
Figure~\ref{fig:latched}(b) presents the signal obtained at the triple point where (1,1), (0,2) and (1,2) are degenerated after a preparation in the (1,1) region in a mixture of spin states. 
The single-shot measurements show the apparition of a step like feature which is the signature of a triplet relaxation at the triple point. 
As a consequence, we can discriminate between singlet and triplet states by looking at the detector level at short time. 
The gate 1 voltage window where these events are observed is determined by the energy separation between singlet and triplet states in the (0,2) charge configuration, which is equal to the valley splitting in the dot 2 (see Fig.~\ref{fig:latched}(a)). 
We measure a valley splitting of $150$~\si{\micro\electronvolt} comparable to what has been obtained in planar MOS devices (data not shown) \cite{yang2013spin}. 
Figure~\ref{fig:fidelity}(a) presents the histogram of the single shot measurement for an integration time of $1\,ms$ for mixed population between singlet and triplet states. 
It clearly shows two Gaussian distributions corresponding to the two spin states. 
The readout visibility, see Fig.~\ref{fig:fidelity}(b) can be as high as 99.3\% giving singlet and triplet readout fidelities of 99.7\%. 
Figure~\ref{fig:fidelity}(c) presents the error rate as a function of the integration time. 
At short times, the fidelity is limited by the signal over noise ratio. 
As the integration time increases the error rate decreases until it reaches a minimum around $1.5$~\si{\milli\second}. 
Above this value the triplet state relaxes during the measurement time as will be developed in the following.

Besides detection errors that degrade readout fidelity, other sources of errors can alter the measurement. 
In order to evaluate the physical errors that can occur, we set the magnetic field to $3$~\si{\tesla} to separate the $\text{T}_-$ ground state and the excited singlet state with an energy much larger than the temperature. 
We then prepare a singlet state in the (0,2) region that we adiabatically transfer to (1,1). 
We check that this transfer conserves the spin by pulsing to the measurement point where we find 96.8\% singlet population. 
Once the transfer is achieved, we let the singlet (1,1) relax to the $\text{T}_-$ ground state followed by a pulse to the measurement point. 
Figure~\ref{fig:fidelity}(d) presents such measurement which corresponds to a $T_1$ measurement. 
It shows that the final population of singlet is not null while the Boltzmann distribution gives almost 100\% triplet. 
The error is likely to occur during the measurement where the triplet can be transferred to the singlet before the measurement is complete. An estimation of this relaxation time compared to the integration time gives a systematic error of 3.5\%.
\begin{figure}[t]%
\includegraphics[width=\columnwidth]{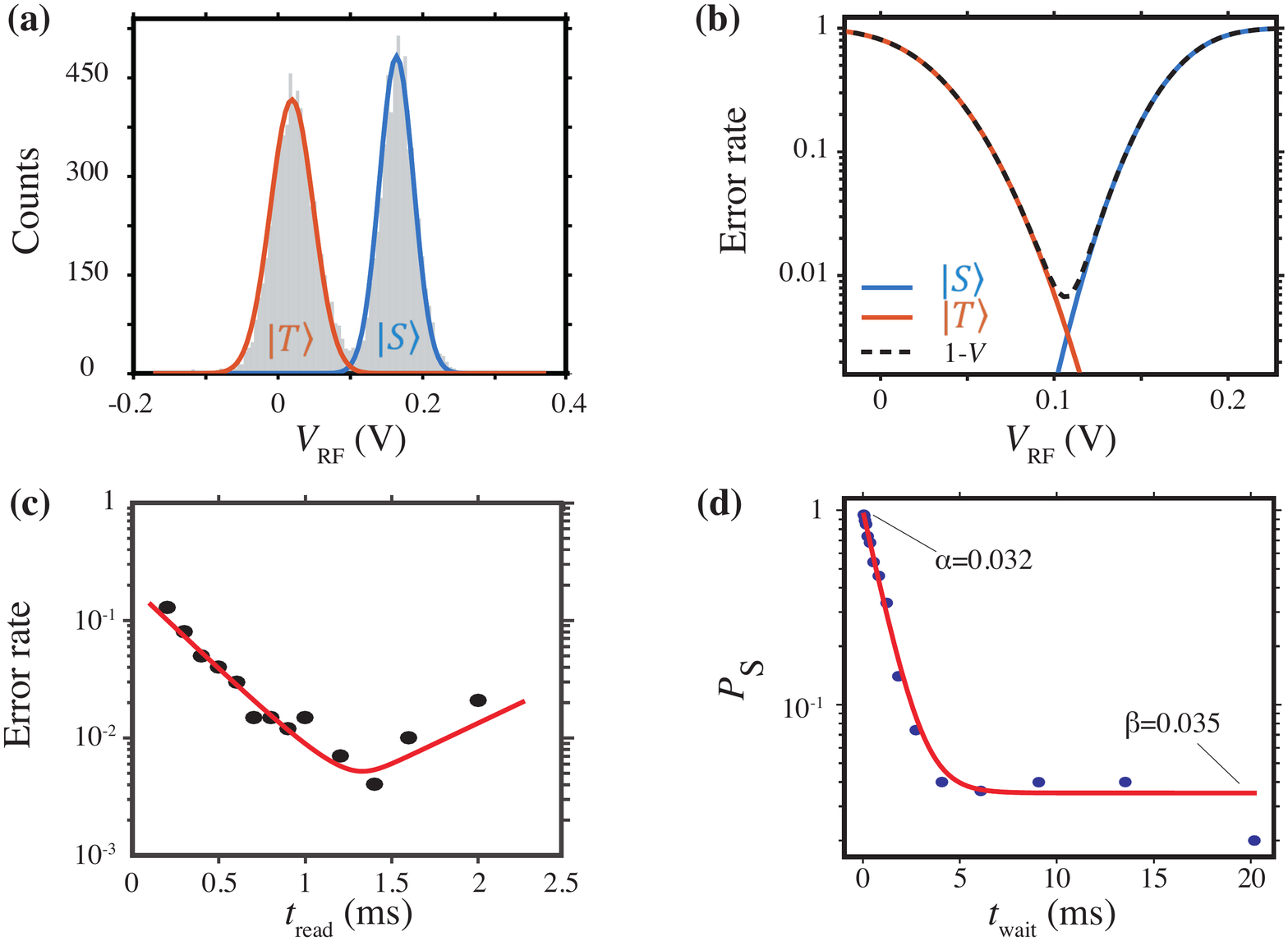}%
\caption{\textbf{Spin readout error analysis}. 
(a) Histogram of the signal distribution. The two bell curves correspond to the two spin states triplet and singlet and are fitted with Gaussian distributions. 
(b) The orange and blue solid lines correspond respectively to triplet and singlet readout fidelity. The dashed line corresponds to 1-$V$ where $V$ is the visibility. At the point of maximum visibility, the fidelity reads $99.7\%$ for both spin state. (c) The fidelity is plotted as function of measurement time ($t_{\text{read}}$). At short time, the fidelity is degraded because of the noise in the amplification chain. The fidelity reaches a maximum around 1ms and then drops down for longer integration time due to relaxation of the spin during the measurement. 
(d) \textit{P} represents the probability to measure a singlet state at M and is plotted as a function of the time spent at W ($t_{\text{wait}}$). The red solid line corresponds to an exponential fit. This relaxation curve allows to extract the initialization and $T_1$ errors which are respectively $\alpha = 3.2\%$ and $\beta=3.5\%$.
}
\label{fig:fidelity}%
\end{figure}

\begin{figure}[t]%
\includegraphics[width=\columnwidth]{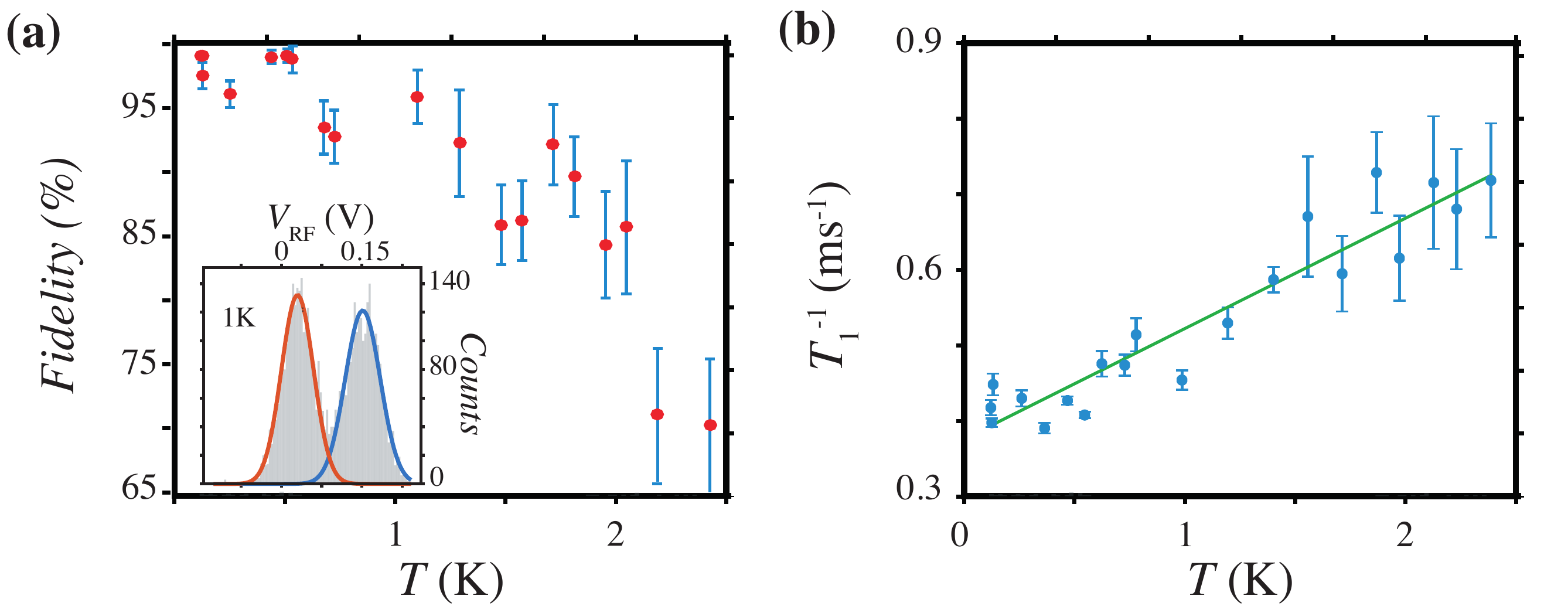}%
\caption{\textbf{Temperature dependence of the readout fidelity and relaxation rate}. 
(a) The readout fidelity stays almost constant below $500$~\si{\milli\kelvin} and drops drastically after $1$~\si{\kelvin} due to thermal broadening of the detector. The inset shows the spin detection distribution at $1$~\si{\kelvin} with a readout fidelity of 94\%. 
(b) The relaxation rate is measured as a function of temperature and presents a linear dependence as expected for a direct relaxation mechanism.}
\label{fig:temp}%
\end{figure}

An interesting feature of the present readout procedure is the possibility to work at relatively high temperature. 
Indeed, in contrast with the energy selective tunneling readout, the latched PSB mechanism is based on spin-dependent tunneling with no constrain on the ratio between Zeeman and thermal energies. 
We investigate the temperature dependence of the readout fidelity up to $2$\si{\kelvin} at $3$~\si{\tesla} in Fig.~\ref{fig:temp}(a). 
We can keep high-fidelity readout up to $500$~\si{\milli\kelvin} and the fidelity then decreases with increasing temperature. 
Whereas the width of the Gaussian distribution associated to each spin state are determined by the cryogenic amplifier noise and are therefore insensitive to the temperature of the electrons, the separation between the two Gaussian distributions is decreasing with temperature due the detector Coulomb peak broadening. 
The cross-over is determined by the coupling between the detector and the reservoir that is estimated to $50$~\si{\micro\electronvolt}. 
Moreover, we have measured the relaxation time as a function of temperature (see Fig.~\ref{fig:temp}(b)), we observed a linear decrease as a function of the temperature with only a reduction by a factor 2 at $2$\si{\kelvin}. 



\section{Discussion}
The demonstrated fidelity of the spin readout is limited by two factors: the detector/qubit capacitive coupling and the noise of the cryogenic amplifier. First, increasing the coupling can be obtained by optimizing the design with respect to the parasitic capacitances. As far as noise is concerned, our homemade cryogenic low noise amplifier has a measured noise temperature of $70$~\si{\kelvin}. This is more than two orders of magnitude larger than the noise temperature obtained with state-of-the-art superconducting amplifiers \citep{macklin2015near}. In total, we could envision to keep the fidelity as high as demonstrated in the present manuscript and to reduce the integration time below $10$~\si{\micro\second}. All the qubit operations would then be in the \si{\micro\second}-range, a speed that makes large-scale computation viable in terms of computational run-time.
Besides the demonstration of the high-fidelity readout, we have shown that our technique is robust up to $0.5$~\si{\kelvin}. Being able to readout the spin of an electron at higher temperature offers perspectives for large-scale integration and for building an efficient quantum/classical interface. Indeed, large scale integration requires a complex control hardware which dissipates heat and therefore requires cooling power to keep the system cold. We can expect from modern cryogenics to obtain more than $100$~\si{\milli\watt} cooling power at $0.5$~\si{\kelvin}. Therefore, we aim to develop a complex classical control system on the same chip as the quantum hardware and to use non-equilibrium manipulation schemes possible for spin qubits considering their long relaxation time at $0.5$~\si{\kelvin}, as shown by the present letter and recent experiments \cite{PhysRevLett.121.076801}.


\section{Conclusion}
In the present manuscript, we have demonstrated high fidelity and robust spin readout using RF-gate reflectometry in a CMOS double dot device. Our procedure is compatible with a scalable architecture where helper dots connected to a single reservoir are locally coupled to each electron spin qubit of the 2D array \cite{usineagaz}. It is worth mentioning that having local reservoirs could greatly simplify the electron loading and qubit initialization procedures of the 2D electron spin qubit array. Perspectives to engineer a \si{\micro\second}-timescale and multiplexed high fidelity readout with an optimized RF set-up would put electron spin qubit in a favorable position to perform quantum information processing.

\section{Methods}
\textbf{Materials and set-up.} The device, sketched in Fig.~1(a) is fabricated from an SOI substrate composed of a $145$~\si{\nano\metre} buried oxide layer and a $11$~\si{\nano\metre} thick silicon layer. 
The thin silicon film is patterned to create a $200$~\si{\nano\metre} long and $30$~\si{\nano\metre} wide nanowire by means of e-beam lithography. 
Three 30nm wide wrap-around top gates are defined using a SiO$_2$($2.5$~\si{\nano\metre})/HfO$_2$($1.9$~\si{\nano\metre}) stack for the gate dielectric followed by TiN($5$~\si{\nano\metre})/poly-Si($50$~\si{\nano\metre}) as the top gate material. 
The source and drain are self aligned and formed by phosphorous ion implantation and annealing after the deposition of $20$~\si{\nano\metre} long Si$_3$N$_4$ spacers. 
The device is anchored to the cold finger, which is in turn mechanically attached to the mixing chamber of a homemade dilution refrigerator with a base temperature of $80$~\si{\milli\kelvin}. It is placed at the center of a superconducting solenoid generating the static out-of-plane magnetic field. Quantum dots are defined and controlled by the application of voltages on gates deposited on the surface of the crystal. Homemade electronics ensure fast changes of both chemical potentials and tunnel couplings with voltage pulse rise times approaching $100$~\si{\nano\second} and refreshed every $16$~\si{\micro\second}. 

The tank circuit is composed of a surface mounted inductance ($820$~\si{\nano\henry}), a parasitic capacitance to ground ($0.75$~\si{\pico\farad}) and the device capacitance between gate 1 and the device channel. 
RF-reflectometry is performed close to resonance frequency (234 MHz), with the input power set to $-95$~\si{\dBm} and the reflected signal amplified by a low noise cryogenic amplifier anchored at the $4$~\si{\kelvin} stage. 
The signal is further amplified and demodulated at room temperature as shown in Fig.~1(b). 
Moreover, a switch is used to turn on the RF excitation only during the measurement sequence.

\section{Acknowledgment}
We acknowledge technical support from the workshops at the Institut N\'eel, in particular from P. Perrier, H. Rodenas, E. Eyraud, D. Lepoittevin, C. Hoarau and C. Guttin. 
M.U. acknowledges the support from Marie Sklodowska Curie fellowship within the horizon 2020 european program. This work is supported by the Agence Nationale de la Recherche through the projects ANR-16-ACHN-0029 and ANR-15-IDEX-02.

\bibliography{biblio}

\begin{thebibliography}{31}%
\makeatletter
\providecommand \@ifxundefined [1]{%
 \@ifx{#1\undefined}
}%
\providecommand \@ifnum [1]{%
 \ifnum #1\expandafter \@firstoftwo
 \else \expandafter \@secondoftwo
 \fi
}%
\providecommand \@ifx [1]{%
 \ifx #1\expandafter \@firstoftwo
 \else \expandafter \@secondoftwo
 \fi
}%
\providecommand \natexlab [1]{#1}%
\providecommand \enquote  [1]{``#1''}%
\providecommand \bibnamefont  [1]{#1}%
\providecommand \bibfnamefont [1]{#1}%
\providecommand \citenamefont [1]{#1}%
\providecommand \href@noop [0]{\@secondoftwo}%
\providecommand \href [0]{\begingroup \@sanitize@url \@href}%
\providecommand \@href[1]{\@@startlink{#1}\@@href}%
\providecommand \@@href[1]{\endgroup#1\@@endlink}%
\providecommand \@sanitize@url [0]{\catcode `\\12\catcode `\$12\catcode
  `\&12\catcode `\#12\catcode `\^12\catcode `\_12\catcode `\%12\relax}%
\providecommand \@@startlink[1]{}%
\providecommand \@@endlink[0]{}%
\providecommand \url  [0]{\begingroup\@sanitize@url \@url }%
\providecommand \@url [1]{\endgroup\@href {#1}{\urlprefix }}%
\providecommand \urlprefix  [0]{URL }%
\providecommand \Eprint [0]{\href }%
\providecommand \doibase [0]{http://dx.doi.org/}%
\providecommand \selectlanguage [0]{\@gobble}%
\providecommand \bibinfo  [0]{\@secondoftwo}%
\providecommand \bibfield  [0]{\@secondoftwo}%
\providecommand \translation [1]{[#1]}%
\providecommand \BibitemOpen [0]{}%
\providecommand \bibitemStop [0]{}%
\providecommand \bibitemNoStop [0]{.\EOS\space}%
\providecommand \EOS [0]{\spacefactor3000\relax}%
\providecommand \BibitemShut  [1]{\csname bibitem#1\endcsname}%
\let\auto@bib@innerbib\@empty
\bibitem [{\citenamefont {Fowler}(2011)}]{Fowler2011}%
  \BibitemOpen
  \bibfield  {author} {\bibinfo {author} {\bibfnamefont {A.~G.}\ \bibnamefont
  {Fowler}},\ }\href@noop {} {\bibfield  {journal} {\bibinfo  {journal} {Phys.
  Rev. A}\ }\textbf {\bibinfo {volume} {83}},\ \bibinfo {pages} {042310}
  (\bibinfo {year} {2011})}\BibitemShut {NoStop}%
\bibitem [{\citenamefont {Maurand}\ \emph {et~al.}(2016)\citenamefont
  {Maurand}, \citenamefont {Jehl}, \citenamefont {Kotekar-Patil}, \citenamefont
  {Corna}, \citenamefont {Bohuslavskyi}, \citenamefont {Lavi{\'{e}}ville},
  \citenamefont {Hutin}, \citenamefont {Barraud}, \citenamefont {Vinet},
  \citenamefont {Sanquer},\ and\ \citenamefont {{De
  Franceschi}}}]{Maurand2016}%
  \BibitemOpen
  \bibfield  {author} {\bibinfo {author} {\bibfnamefont {R.}~\bibnamefont
  {Maurand}}, \bibinfo {author} {\bibfnamefont {X.}~\bibnamefont {Jehl}},
  \bibinfo {author} {\bibfnamefont {D.}~\bibnamefont {Kotekar-Patil}}, \bibinfo
  {author} {\bibfnamefont {A.}~\bibnamefont {Corna}}, \bibinfo {author}
  {\bibfnamefont {H.}~\bibnamefont {Bohuslavskyi}}, \bibinfo {author}
  {\bibfnamefont {R.}~\bibnamefont {Lavi{\'{e}}ville}}, \bibinfo {author}
  {\bibfnamefont {L.}~\bibnamefont {Hutin}}, \bibinfo {author} {\bibfnamefont
  {S.}~\bibnamefont {Barraud}}, \bibinfo {author} {\bibfnamefont
  {M.}~\bibnamefont {Vinet}}, \bibinfo {author} {\bibfnamefont
  {M.}~\bibnamefont {Sanquer}}, \ and\ \bibinfo {author} {\bibfnamefont
  {S.}~\bibnamefont {{De Franceschi}}},\ }\href@noop {} {\bibfield  {journal}
  {\bibinfo  {journal} {Nat. Commun.}\ }\textbf {\bibinfo {volume} {7}},\
  \bibinfo {pages} {13575} (\bibinfo {year} {2016})}\BibitemShut {NoStop}%
\bibitem [{\citenamefont {Veldhorst}\ \emph {et~al.}(2015)\citenamefont
  {Veldhorst}, \citenamefont {Yang}, \citenamefont {Hwang}, \citenamefont
  {Huang}, \citenamefont {Dehollain}, \citenamefont {Muhonen}, \citenamefont
  {Simmons}, \citenamefont {Laucht}, \citenamefont {Hudson}, \citenamefont
  {Itoh}, \citenamefont {Morello},\ and\ \citenamefont
  {Dzurak}}]{Veldhorst2015}%
  \BibitemOpen
  \bibfield  {author} {\bibinfo {author} {\bibfnamefont {M.}~\bibnamefont
  {Veldhorst}}, \bibinfo {author} {\bibfnamefont {C.~H.}\ \bibnamefont {Yang}},
  \bibinfo {author} {\bibfnamefont {J.~C.~C.}\ \bibnamefont {Hwang}}, \bibinfo
  {author} {\bibfnamefont {W.}~\bibnamefont {Huang}}, \bibinfo {author}
  {\bibfnamefont {J.~P.}\ \bibnamefont {Dehollain}}, \bibinfo {author}
  {\bibfnamefont {J.~T.}\ \bibnamefont {Muhonen}}, \bibinfo {author}
  {\bibfnamefont {S.}~\bibnamefont {Simmons}}, \bibinfo {author} {\bibfnamefont
  {A.}~\bibnamefont {Laucht}}, \bibinfo {author} {\bibfnamefont {F.~E.}\
  \bibnamefont {Hudson}}, \bibinfo {author} {\bibfnamefont {K.~M.}\
  \bibnamefont {Itoh}}, \bibinfo {author} {\bibfnamefont {A.}~\bibnamefont
  {Morello}}, \ and\ \bibinfo {author} {\bibfnamefont {A.~S.}\ \bibnamefont
  {Dzurak}},\ }\href@noop {} {\bibfield  {journal} {\bibinfo  {journal}
  {Nature}\ }\textbf {\bibinfo {volume} {526}},\ \bibinfo {pages} {410}
  (\bibinfo {year} {2015})}\BibitemShut {NoStop}%
\bibitem [{\citenamefont {Zajac}\ \emph {et~al.}(2018)\citenamefont {Zajac},
  \citenamefont {Sigillito}, \citenamefont {Russ}, \citenamefont {Borjans},
  \citenamefont {Taylor}, \citenamefont {Burkard},\ and\ \citenamefont
  {Petta}}]{Zajac2018}%
  \BibitemOpen
  \bibfield  {author} {\bibinfo {author} {\bibfnamefont {D.~M.}\ \bibnamefont
  {Zajac}}, \bibinfo {author} {\bibfnamefont {A.~J.}\ \bibnamefont
  {Sigillito}}, \bibinfo {author} {\bibfnamefont {M.}~\bibnamefont {Russ}},
  \bibinfo {author} {\bibfnamefont {F.}~\bibnamefont {Borjans}}, \bibinfo
  {author} {\bibfnamefont {J.~M.}\ \bibnamefont {Taylor}}, \bibinfo {author}
  {\bibfnamefont {G.}~\bibnamefont {Burkard}}, \ and\ \bibinfo {author}
  {\bibfnamefont {J.~R.}\ \bibnamefont {Petta}},\ }\href@noop {} {\bibfield
  {journal} {\bibinfo  {journal} {Science}\ }\textbf {\bibinfo {volume}
  {359}},\ \bibinfo {pages} {439} (\bibinfo {year} {2018})}\BibitemShut
  {NoStop}%
\bibitem [{\citenamefont {Watson}\ \emph {et~al.}(2018)\citenamefont {Watson},
  \citenamefont {Philips}, \citenamefont {Kawakami}, \citenamefont {Ward},
  \citenamefont {Scarlino}, \citenamefont {Veldhorst}, \citenamefont {Savage},
  \citenamefont {Lagally}, \citenamefont {Friesen}, \citenamefont
  {Coppersmith}, \citenamefont {Erikson},\ and\ \citenamefont
  {Vandersypen}}]{Watson2018}%
  \BibitemOpen
  \bibfield  {author} {\bibinfo {author} {\bibfnamefont {T.~F.}\ \bibnamefont
  {Watson}}, \bibinfo {author} {\bibfnamefont {S.~G.~J.}\ \bibnamefont
  {Philips}}, \bibinfo {author} {\bibfnamefont {E.}~\bibnamefont {Kawakami}},
  \bibinfo {author} {\bibfnamefont {D.~R.}\ \bibnamefont {Ward}}, \bibinfo
  {author} {\bibfnamefont {P.}~\bibnamefont {Scarlino}}, \bibinfo {author}
  {\bibfnamefont {M.}~\bibnamefont {Veldhorst}}, \bibinfo {author}
  {\bibfnamefont {D.~E.}\ \bibnamefont {Savage}}, \bibinfo {author}
  {\bibfnamefont {M.~G.}\ \bibnamefont {Lagally}}, \bibinfo {author}
  {\bibfnamefont {M.}~\bibnamefont {Friesen}}, \bibinfo {author} {\bibfnamefont
  {S.~N.}\ \bibnamefont {Coppersmith}}, \bibinfo {author} {\bibfnamefont
  {M.~A.}\ \bibnamefont {Erikson}}, \ and\ \bibinfo {author} {\bibfnamefont
  {L.~M.~K.}\ \bibnamefont {Vandersypen}},\ }\href@noop {} {\bibfield
  {journal} {\bibinfo  {journal} {Nature}\ }\textbf {\bibinfo {volume} {555}},\
  \bibinfo {pages} {633} (\bibinfo {year} {2018})}\BibitemShut {NoStop}%
\bibitem [{\citenamefont {Yoneda}\ \emph {et~al.}(2018)\citenamefont {Yoneda},
  \citenamefont {Takeda}, \citenamefont {Otsuka}, \citenamefont {Nakajima},
  \citenamefont {Delbecq}, \citenamefont {Allison}, \citenamefont {Honda},
  \citenamefont {Kodera}, \citenamefont {Oda}, \citenamefont {Hoshi} \emph
  {et~al.}}]{yoneda2018quantum}%
  \BibitemOpen
  \bibfield  {author} {\bibinfo {author} {\bibfnamefont {J.}~\bibnamefont
  {Yoneda}}, \bibinfo {author} {\bibfnamefont {K.}~\bibnamefont {Takeda}},
  \bibinfo {author} {\bibfnamefont {T.}~\bibnamefont {Otsuka}}, \bibinfo
  {author} {\bibfnamefont {T.}~\bibnamefont {Nakajima}}, \bibinfo {author}
  {\bibfnamefont {M.~R.}\ \bibnamefont {Delbecq}}, \bibinfo {author}
  {\bibfnamefont {G.}~\bibnamefont {Allison}}, \bibinfo {author} {\bibfnamefont
  {T.}~\bibnamefont {Honda}}, \bibinfo {author} {\bibfnamefont
  {T.}~\bibnamefont {Kodera}}, \bibinfo {author} {\bibfnamefont
  {S.}~\bibnamefont {Oda}}, \bibinfo {author} {\bibfnamefont {Y.}~\bibnamefont
  {Hoshi}},  \emph {et~al.},\ }\href@noop {} {\bibfield  {journal} {\bibinfo
  {journal} {Nature Nanotech.}\ }\textbf {\bibinfo {volume} {13}},\ \bibinfo
  {pages} {102} (\bibinfo {year} {2018})}\BibitemShut {NoStop}%
\bibitem [{\citenamefont {Vandersypen}\ \emph {et~al.}(2017)\citenamefont
  {Vandersypen}, \citenamefont {Bluhm}, \citenamefont {Clarke}, \citenamefont
  {Dzurak}, \citenamefont {Ishihara}, \citenamefont {Morello}, \citenamefont
  {Reilly}, \citenamefont {Schreider},\ and\ \citenamefont
  {Veldhorst}}]{Vandersypen2017}%
  \BibitemOpen
  \bibfield  {author} {\bibinfo {author} {\bibfnamefont {L.~M.~K.}\
  \bibnamefont {Vandersypen}}, \bibinfo {author} {\bibfnamefont
  {H.}~\bibnamefont {Bluhm}}, \bibinfo {author} {\bibfnamefont {J.~S.}\
  \bibnamefont {Clarke}}, \bibinfo {author} {\bibfnamefont {A.~S.}\
  \bibnamefont {Dzurak}}, \bibinfo {author} {\bibfnamefont {J.}~\bibnamefont
  {Ishihara}}, \bibinfo {author} {\bibfnamefont {A.}~\bibnamefont {Morello}},
  \bibinfo {author} {\bibfnamefont {D.~J.}\ \bibnamefont {Reilly}}, \bibinfo
  {author} {\bibfnamefont {L.~R.}\ \bibnamefont {Schreider}}, \ and\ \bibinfo
  {author} {\bibfnamefont {M.}~\bibnamefont {Veldhorst}},\ }\href@noop {}
  {\bibfield  {journal} {\bibinfo  {journal} {npj Quantum Information}\
  }\textbf {\bibinfo {volume} {3}},\ \bibinfo {pages} {34} (\bibinfo {year}
  {2017})}\BibitemShut {NoStop}%
\bibitem [{\citenamefont {Veldhorst}\ \emph {et~al.}(2017)\citenamefont
  {Veldhorst}, \citenamefont {Eenink}, \citenamefont {Yang},\ and\
  \citenamefont {Dzurak}}]{Veldhorst2017}%
  \BibitemOpen
  \bibfield  {author} {\bibinfo {author} {\bibfnamefont {M.}~\bibnamefont
  {Veldhorst}}, \bibinfo {author} {\bibfnamefont {H.~G.~J.}\ \bibnamefont
  {Eenink}}, \bibinfo {author} {\bibfnamefont {C.~H.}\ \bibnamefont {Yang}}, \
  and\ \bibinfo {author} {\bibfnamefont {A.~S.}\ \bibnamefont {Dzurak}},\
  }\href@noop {} {\bibfield  {journal} {\bibinfo  {journal} {Nat. Commun.}\
  }\textbf {\bibinfo {volume} {8}},\ \bibinfo {pages} {1766} (\bibinfo {year}
  {2017})}\BibitemShut {NoStop}%
\bibitem [{\citenamefont {Hutin}\ \emph {et~al.}(2018)\citenamefont {Hutin},
  \citenamefont {De~Franceschi}, \citenamefont {Meunier},\ and\ \citenamefont
  {Vinet}}]{usineagaz}%
  \BibitemOpen
  \bibfield  {author} {\bibinfo {author} {\bibfnamefont {L.}~\bibnamefont
  {Hutin}}, \bibinfo {author} {\bibfnamefont {S.}~\bibnamefont
  {De~Franceschi}}, \bibinfo {author} {\bibfnamefont {T.}~\bibnamefont
  {Meunier}}, \ and\ \bibinfo {author} {\bibfnamefont {M.}~\bibnamefont
  {Vinet}},\ }\href@noop {} {\enquote {\bibinfo {title} {Quantum device with
  spin qubits},}\ } (\bibinfo {year} {2018}),\ \bibinfo {note} {u.S.
  Provisionnal Pat. Ser. No. 15/967778}\BibitemShut {NoStop}%
\bibitem [{\citenamefont {Thalineau}\ \emph {et~al.}(2012)\citenamefont
  {Thalineau}, \citenamefont {Hermelin}, \citenamefont {Wieck}, \citenamefont
  {B{\"a}uerle}, \citenamefont {Saminadayar},\ and\ \citenamefont
  {Meunier}}]{Thalineau2012}%
  \BibitemOpen
  \bibfield  {author} {\bibinfo {author} {\bibfnamefont {R.}~\bibnamefont
  {Thalineau}}, \bibinfo {author} {\bibfnamefont {S.}~\bibnamefont {Hermelin}},
  \bibinfo {author} {\bibfnamefont {A.~D.}\ \bibnamefont {Wieck}}, \bibinfo
  {author} {\bibfnamefont {C.}~\bibnamefont {B{\"a}uerle}}, \bibinfo {author}
  {\bibfnamefont {L.}~\bibnamefont {Saminadayar}}, \ and\ \bibinfo {author}
  {\bibfnamefont {T.}~\bibnamefont {Meunier}},\ }\href@noop {} {\bibfield
  {journal} {\bibinfo  {journal} {Appl. Phys. Lett.}\ }\textbf {\bibinfo
  {volume} {101}},\ \bibinfo {pages} {103102} (\bibinfo {year}
  {2012})}\BibitemShut {NoStop}%
\bibitem [{\citenamefont {Flentje}\ \emph {et~al.}(2017)\citenamefont
  {Flentje}, \citenamefont {Bertrand}, \citenamefont {Mortemousque},
  \citenamefont {Thiney}, \citenamefont {Ludwig}, \citenamefont {Wieck},
  \citenamefont {B{\"a}uerle},\ and\ \citenamefont {Meunier}}]{Flentje2017}%
  \BibitemOpen
  \bibfield  {author} {\bibinfo {author} {\bibfnamefont {H.}~\bibnamefont
  {Flentje}}, \bibinfo {author} {\bibfnamefont {B.}~\bibnamefont {Bertrand}},
  \bibinfo {author} {\bibfnamefont {P.~A.}\ \bibnamefont {Mortemousque}},
  \bibinfo {author} {\bibfnamefont {V.}~\bibnamefont {Thiney}}, \bibinfo
  {author} {\bibfnamefont {A.}~\bibnamefont {Ludwig}}, \bibinfo {author}
  {\bibfnamefont {A.~D.}\ \bibnamefont {Wieck}}, \bibinfo {author}
  {\bibfnamefont {C.}~\bibnamefont {B{\"a}uerle}}, \ and\ \bibinfo {author}
  {\bibfnamefont {T.}~\bibnamefont {Meunier}},\ }\href@noop {} {\bibfield
  {journal} {\bibinfo  {journal} {Appl. Phys. Lett.}\ }\textbf {\bibinfo
  {volume} {110}},\ \bibinfo {pages} {233101} (\bibinfo {year}
  {2017})}\BibitemShut {NoStop}%
\bibitem [{\citenamefont {Mukhopadhyay}\ \emph {et~al.}(2018)\citenamefont
  {Mukhopadhyay}, \citenamefont {Dehollain}, \citenamefont {Reichl},
  \citenamefont {Wegscheider},\ and\ \citenamefont
  {Vandersypen}}]{Mukhopadhyay2018}%
  \BibitemOpen
  \bibfield  {author} {\bibinfo {author} {\bibfnamefont {U.}~\bibnamefont
  {Mukhopadhyay}}, \bibinfo {author} {\bibfnamefont {J.~P.}\ \bibnamefont
  {Dehollain}}, \bibinfo {author} {\bibfnamefont {C.}~\bibnamefont {Reichl}},
  \bibinfo {author} {\bibfnamefont {W.}~\bibnamefont {Wegscheider}}, \ and\
  \bibinfo {author} {\bibfnamefont {L.~M.~K.}\ \bibnamefont {Vandersypen}},\
  }\href@noop {} {\bibfield  {journal} {\bibinfo  {journal} {Appl. Phys.
  Lett.}\ }\textbf {\bibinfo {volume} {112}},\ \bibinfo {pages} {183505}
  (\bibinfo {year} {2018})}\BibitemShut {NoStop}%
\bibitem [{\citenamefont {Mortemousque}\ \emph {et~al.}()\citenamefont
  {Mortemousque}, \citenamefont {Chanrion}, \citenamefont {Jadot},
  \citenamefont {Flentje}, \citenamefont {Ludwig}, \citenamefont {Wieck},
  \citenamefont {Urdampilleta}, \citenamefont {Bauerle},\ and\ \citenamefont
  {Meunier}}]{Mortemousque2018}%
  \BibitemOpen
  \bibfield  {author} {\bibinfo {author} {\bibfnamefont {P.~A.}\ \bibnamefont
  {Mortemousque}}, \bibinfo {author} {\bibfnamefont {E.}~\bibnamefont
  {Chanrion}}, \bibinfo {author} {\bibfnamefont {B.}~\bibnamefont {Jadot}},
  \bibinfo {author} {\bibfnamefont {H.}~\bibnamefont {Flentje}}, \bibinfo
  {author} {\bibfnamefont {A.}~\bibnamefont {Ludwig}}, \bibinfo {author}
  {\bibfnamefont {A.~D.}\ \bibnamefont {Wieck}}, \bibinfo {author}
  {\bibfnamefont {M.}~\bibnamefont {Urdampilleta}}, \bibinfo {author}
  {\bibfnamefont {C.}~\bibnamefont {Bauerle}}, \ and\ \bibinfo {author}
  {\bibfnamefont {T.}~\bibnamefont {Meunier}},\ }\href@noop {} {\bibinfo
  {journal} {arXiv:1808.06180}\ }\BibitemShut {NoStop}%
\bibitem [{\citenamefont {Li}\ \emph {et~al.}(2018)\citenamefont {Li},
  \citenamefont {Petit}, \citenamefont {Franke}, \citenamefont {Dehollain},
  \citenamefont {Helsen}, \citenamefont {Steudtner}, \citenamefont {Thomas},
  \citenamefont {Yoscovits}, \citenamefont {Singh}, \citenamefont {Wehner}
  \emph {et~al.}}]{li2018crossbar}%
  \BibitemOpen
\bibfield  {journal} {  }\bibfield  {author} {\bibinfo {author} {\bibfnamefont
  {R.}~\bibnamefont {Li}}, \bibinfo {author} {\bibfnamefont {L.}~\bibnamefont
  {Petit}}, \bibinfo {author} {\bibfnamefont {D.~P.}\ \bibnamefont {Franke}},
  \bibinfo {author} {\bibfnamefont {J.~P.}\ \bibnamefont {Dehollain}}, \bibinfo
  {author} {\bibfnamefont {J.}~\bibnamefont {Helsen}}, \bibinfo {author}
  {\bibfnamefont {M.}~\bibnamefont {Steudtner}}, \bibinfo {author}
  {\bibfnamefont {N.~K.}\ \bibnamefont {Thomas}}, \bibinfo {author}
  {\bibfnamefont {Z.~R.}\ \bibnamefont {Yoscovits}}, \bibinfo {author}
  {\bibfnamefont {K.~J.}\ \bibnamefont {Singh}}, \bibinfo {author}
  {\bibfnamefont {S.}~\bibnamefont {Wehner}},  \emph {et~al.},\ }\href@noop {}
  {\bibfield  {journal} {\bibinfo  {journal} {Sci. Adv.}\ }\textbf {\bibinfo
  {volume} {4}},\ \bibinfo {pages} {eaar3960} (\bibinfo {year}
  {2018})}\BibitemShut {NoStop}%
\bibitem [{\citenamefont {Elzerman}\ \emph {et~al.}(2004)\citenamefont
  {Elzerman}, \citenamefont {Hanson}, \citenamefont {{Willems van Beveren}},
  \citenamefont {Witkamp}, \citenamefont {Vandersypen},\ and\ \citenamefont
  {Kouwenhoven}}]{Elzerman2004}%
  \BibitemOpen
  \bibfield  {author} {\bibinfo {author} {\bibfnamefont {J.~M.}\ \bibnamefont
  {Elzerman}}, \bibinfo {author} {\bibfnamefont {R.}~\bibnamefont {Hanson}},
  \bibinfo {author} {\bibfnamefont {L.~H.}\ \bibnamefont {{Willems van
  Beveren}}}, \bibinfo {author} {\bibfnamefont {B.}~\bibnamefont {Witkamp}},
  \bibinfo {author} {\bibfnamefont {L.~M.~K.}\ \bibnamefont {Vandersypen}}, \
  and\ \bibinfo {author} {\bibfnamefont {L.~P.}\ \bibnamefont {Kouwenhoven}},\
  }\href@noop {} {\bibfield  {journal} {\bibinfo  {journal} {Nature}\ }\textbf
  {\bibinfo {volume} {430}},\ \bibinfo {pages} {431} (\bibinfo {year}
  {2004})}\BibitemShut {NoStop}%
\bibitem [{\citenamefont {Ono}\ \emph {et~al.}(2002)\citenamefont {Ono},
  \citenamefont {Austing}, \citenamefont {Tokura},\ and\ \citenamefont
  {Tarucha}}]{ono2002current}%
  \BibitemOpen
  \bibfield  {author} {\bibinfo {author} {\bibfnamefont {K.}~\bibnamefont
  {Ono}}, \bibinfo {author} {\bibfnamefont {D.}~\bibnamefont {Austing}},
  \bibinfo {author} {\bibfnamefont {Y.}~\bibnamefont {Tokura}}, \ and\ \bibinfo
  {author} {\bibfnamefont {S.}~\bibnamefont {Tarucha}},\ }\href@noop {}
  {\bibfield  {journal} {\bibinfo  {journal} {Science}\ }\textbf {\bibinfo
  {volume} {297}},\ \bibinfo {pages} {1313} (\bibinfo {year}
  {2002})}\BibitemShut {NoStop}%
\bibitem [{\citenamefont {Barthel}\ \emph {et~al.}(2009)\citenamefont
  {Barthel}, \citenamefont {Reilly}, \citenamefont {Marcus}, \citenamefont
  {Hanson},\ and\ \citenamefont {Gossard}}]{PhysRevLett.103.160503}%
  \BibitemOpen
  \bibfield  {author} {\bibinfo {author} {\bibfnamefont {C.}~\bibnamefont
  {Barthel}}, \bibinfo {author} {\bibfnamefont {D.~J.}\ \bibnamefont {Reilly}},
  \bibinfo {author} {\bibfnamefont {C.~M.}\ \bibnamefont {Marcus}}, \bibinfo
  {author} {\bibfnamefont {M.~P.}\ \bibnamefont {Hanson}}, \ and\ \bibinfo
  {author} {\bibfnamefont {A.~C.}\ \bibnamefont {Gossard}},\ }\href@noop {}
  {\bibfield  {journal} {\bibinfo  {journal} {Phys. Rev. Lett.}\ }\textbf
  {\bibinfo {volume} {103}},\ \bibinfo {pages} {160503} (\bibinfo {year}
  {2009})}\BibitemShut {NoStop}%
\bibitem [{\citenamefont {Neeley}\ \emph {et~al.}(2009)\citenamefont {Neeley},
  \citenamefont {Ansmann}, \citenamefont {Bialczak}, \citenamefont {Hofheinz},
  \citenamefont {Lucero}, \citenamefont {O'Connell}, \citenamefont {Sank},
  \citenamefont {Wang}, \citenamefont {Wenner}, \citenamefont {Cleland},
  \citenamefont {Geller},\ and\ \citenamefont {Martinis}}]{Neeley2009}%
  \BibitemOpen
  \bibfield  {author} {\bibinfo {author} {\bibfnamefont {M.}~\bibnamefont
  {Neeley}}, \bibinfo {author} {\bibfnamefont {M.}~\bibnamefont {Ansmann}},
  \bibinfo {author} {\bibfnamefont {R.~C.}\ \bibnamefont {Bialczak}}, \bibinfo
  {author} {\bibfnamefont {M.}~\bibnamefont {Hofheinz}}, \bibinfo {author}
  {\bibfnamefont {E.}~\bibnamefont {Lucero}}, \bibinfo {author} {\bibfnamefont
  {A.~D.}\ \bibnamefont {O'Connell}}, \bibinfo {author} {\bibfnamefont
  {D.}~\bibnamefont {Sank}}, \bibinfo {author} {\bibfnamefont {H.}~\bibnamefont
  {Wang}}, \bibinfo {author} {\bibfnamefont {J.}~\bibnamefont {Wenner}},
  \bibinfo {author} {\bibfnamefont {A.~N.}\ \bibnamefont {Cleland}}, \bibinfo
  {author} {\bibfnamefont {M.~R.}\ \bibnamefont {Geller}}, \ and\ \bibinfo
  {author} {\bibfnamefont {J.~M.}\ \bibnamefont {Martinis}},\ }\href@noop {}
  {\bibfield  {journal} {\bibinfo  {journal} {Science}\ }\textbf {\bibinfo
  {volume} {325}},\ \bibinfo {pages} {722} (\bibinfo {year}
  {2009})}\BibitemShut {NoStop}%
\bibitem [{\citenamefont {Petersson}\ \emph {et~al.}(2010)\citenamefont
  {Petersson}, \citenamefont {Smith}, \citenamefont {Anderson}, \citenamefont
  {Atkinson}, \citenamefont {Jones},\ and\ \citenamefont
  {Ritchie}}]{petersson2010charge}%
  \BibitemOpen
  \bibfield  {author} {\bibinfo {author} {\bibfnamefont {K.}~\bibnamefont
  {Petersson}}, \bibinfo {author} {\bibfnamefont {C.}~\bibnamefont {Smith}},
  \bibinfo {author} {\bibfnamefont {D.}~\bibnamefont {Anderson}}, \bibinfo
  {author} {\bibfnamefont {P.}~\bibnamefont {Atkinson}}, \bibinfo {author}
  {\bibfnamefont {G.}~\bibnamefont {Jones}}, \ and\ \bibinfo {author}
  {\bibfnamefont {D.}~\bibnamefont {Ritchie}},\ }\href@noop {} {\bibfield
  {journal} {\bibinfo  {journal} {Nano Letters}\ }\textbf {\bibinfo {volume}
  {10}},\ \bibinfo {pages} {2789} (\bibinfo {year} {2010})}\BibitemShut
  {NoStop}%
\bibitem [{\citenamefont {Fogarty}\ \emph {et~al.}()\citenamefont {Fogarty},
  \citenamefont {Chan}, \citenamefont {Hensen}, \citenamefont {Huang},
  \citenamefont {Tanttu}, \citenamefont {Yang}, \citenamefont {Laucht},
  \citenamefont {Veldhorst}, \citenamefont {Hudson}, \citenamefont {Itoh},
  \citenamefont {Culcer}, \citenamefont {Morello},\ and\ \citenamefont
  {Dzurak}}]{fogarty2017integrated}%
  \BibitemOpen
  \bibfield  {author} {\bibinfo {author} {\bibfnamefont {M.}~\bibnamefont
  {Fogarty}}, \bibinfo {author} {\bibfnamefont {K.}~\bibnamefont {Chan}},
  \bibinfo {author} {\bibfnamefont {B.}~\bibnamefont {Hensen}}, \bibinfo
  {author} {\bibfnamefont {W.}~\bibnamefont {Huang}}, \bibinfo {author}
  {\bibfnamefont {T.}~\bibnamefont {Tanttu}}, \bibinfo {author} {\bibfnamefont
  {C.}~\bibnamefont {Yang}}, \bibinfo {author} {\bibfnamefont {A.}~\bibnamefont
  {Laucht}}, \bibinfo {author} {\bibfnamefont {M.}~\bibnamefont {Veldhorst}},
  \bibinfo {author} {\bibfnamefont {F.}~\bibnamefont {Hudson}}, \bibinfo
  {author} {\bibfnamefont {K.}~\bibnamefont {Itoh}}, \bibinfo {author}
  {\bibfnamefont {D.}~\bibnamefont {Culcer}}, \bibinfo {author} {\bibfnamefont
  {A.}~\bibnamefont {Morello}}, \ and\ \bibinfo {author} {\bibfnamefont
  {A.~S.}\ \bibnamefont {Dzurak}},\ }\href@noop {} {\bibinfo  {journal}
  {arXiv:1708.03445}\ }\BibitemShut {NoStop}%
\bibitem [{\citenamefont {Harvey-Collard}\ \emph {et~al.}(2018)\citenamefont
  {Harvey-Collard}, \citenamefont {D'Anjou}, \citenamefont {Rudolph},
  \citenamefont {Jacobson}, \citenamefont {Dominguez}, \citenamefont
  {Ten~Eyck}, \citenamefont {Wendt}, \citenamefont {Pluym}, \citenamefont
  {Lilly}, \citenamefont {Coish}, \citenamefont {Pioro-Ladri\`ere},\ and\
  \citenamefont {Carroll}}]{Harvey}%
  \BibitemOpen
\bibfield  {journal} {  }\bibfield  {author} {\bibinfo {author} {\bibfnamefont
  {P.}~\bibnamefont {Harvey-Collard}}, \bibinfo {author} {\bibfnamefont
  {B.}~\bibnamefont {D'Anjou}}, \bibinfo {author} {\bibfnamefont
  {M.}~\bibnamefont {Rudolph}}, \bibinfo {author} {\bibfnamefont {N.~T.}\
  \bibnamefont {Jacobson}}, \bibinfo {author} {\bibfnamefont {J.}~\bibnamefont
  {Dominguez}}, \bibinfo {author} {\bibfnamefont {G.~A.}\ \bibnamefont
  {Ten~Eyck}}, \bibinfo {author} {\bibfnamefont {J.~R.}\ \bibnamefont {Wendt}},
  \bibinfo {author} {\bibfnamefont {T.}~\bibnamefont {Pluym}}, \bibinfo
  {author} {\bibfnamefont {M.~P.}\ \bibnamefont {Lilly}}, \bibinfo {author}
  {\bibfnamefont {W.~A.}\ \bibnamefont {Coish}}, \bibinfo {author}
  {\bibfnamefont {M.}~\bibnamefont {Pioro-Ladri\`ere}}, \ and\ \bibinfo
  {author} {\bibfnamefont {M.~S.}\ \bibnamefont {Carroll}},\ }\href@noop {}
  {\bibfield  {journal} {\bibinfo  {journal} {Phys. Rev. X}\ }\textbf {\bibinfo
  {volume} {8}},\ \bibinfo {pages} {021046} (\bibinfo {year}
  {2018})}\BibitemShut {NoStop}%
\bibitem [{\citenamefont {Gonzalez-Zalba}\ \emph {et~al.}(2015)\citenamefont
  {Gonzalez-Zalba}, \citenamefont {Barraud}, \citenamefont {Ferguson},\ and\
  \citenamefont {Betz}}]{Gonzalez-Zalba2015}%
  \BibitemOpen
  \bibfield  {author} {\bibinfo {author} {\bibfnamefont {M.~F.}\ \bibnamefont
  {Gonzalez-Zalba}}, \bibinfo {author} {\bibfnamefont {S.}~\bibnamefont
  {Barraud}}, \bibinfo {author} {\bibfnamefont {A.~J.}\ \bibnamefont
  {Ferguson}}, \ and\ \bibinfo {author} {\bibfnamefont {A.~C.}\ \bibnamefont
  {Betz}},\ }\href@noop {} {\bibfield  {journal} {\bibinfo  {journal} {Nat.
  Commun.}\ }\textbf {\bibinfo {volume} {6}},\ \bibinfo {pages} {6084}
  (\bibinfo {year} {2015})}\BibitemShut {NoStop}%
\bibitem [{\citenamefont {Urdampilleta}\ \emph {et~al.}(2015)\citenamefont
  {Urdampilleta}, \citenamefont {Chatterjee}, \citenamefont {Lo}, \citenamefont
  {Kobayashi}, \citenamefont {Mansir}, \citenamefont {Barraud}, \citenamefont
  {Betz}, \citenamefont {Rogge}, \citenamefont {Gonzalez-Zalba},\ and\
  \citenamefont {Morton}}]{Urdampilleta2015a}%
  \BibitemOpen
  \bibfield  {author} {\bibinfo {author} {\bibfnamefont {M.}~\bibnamefont
  {Urdampilleta}}, \bibinfo {author} {\bibfnamefont {A.}~\bibnamefont
  {Chatterjee}}, \bibinfo {author} {\bibfnamefont {C.~C.}\ \bibnamefont {Lo}},
  \bibinfo {author} {\bibfnamefont {T.}~\bibnamefont {Kobayashi}}, \bibinfo
  {author} {\bibfnamefont {J.}~\bibnamefont {Mansir}}, \bibinfo {author}
  {\bibfnamefont {S.}~\bibnamefont {Barraud}}, \bibinfo {author} {\bibfnamefont
  {A.~C.}\ \bibnamefont {Betz}}, \bibinfo {author} {\bibfnamefont
  {S.}~\bibnamefont {Rogge}}, \bibinfo {author} {\bibfnamefont {M.~F.}\
  \bibnamefont {Gonzalez-Zalba}}, \ and\ \bibinfo {author} {\bibfnamefont
  {J.~J.~L.}\ \bibnamefont {Morton}},\ }\href@noop {} {\bibfield  {journal}
  {\bibinfo  {journal} {Phys. Rev. X}\ }\textbf {\bibinfo {volume} {5}},\
  \bibinfo {pages} {031024} (\bibinfo {year} {2015})}\BibitemShut {NoStop}%
\bibitem [{\citenamefont {Hile}\ \emph {et~al.}(2015)\citenamefont {Hile},
  \citenamefont {House}, \citenamefont {Peretz}, \citenamefont {Verduijn},
  \citenamefont {Widmann}, \citenamefont {Kobayashi}, \citenamefont {Rogge},\
  and\ \citenamefont {Simmons}}]{hile2015radio}%
  \BibitemOpen
  \bibfield  {author} {\bibinfo {author} {\bibfnamefont {S.~J.}\ \bibnamefont
  {Hile}}, \bibinfo {author} {\bibfnamefont {M.~G.}\ \bibnamefont {House}},
  \bibinfo {author} {\bibfnamefont {E.}~\bibnamefont {Peretz}}, \bibinfo
  {author} {\bibfnamefont {J.}~\bibnamefont {Verduijn}}, \bibinfo {author}
  {\bibfnamefont {D.}~\bibnamefont {Widmann}}, \bibinfo {author} {\bibfnamefont
  {T.}~\bibnamefont {Kobayashi}}, \bibinfo {author} {\bibfnamefont
  {S.}~\bibnamefont {Rogge}}, \ and\ \bibinfo {author} {\bibfnamefont {M.~Y.}\
  \bibnamefont {Simmons}},\ }\href@noop {} {\bibfield  {journal} {\bibinfo
  {journal} {Appl. Phys. Lett.}\ }\textbf {\bibinfo {volume} {107}},\ \bibinfo
  {pages} {093504} (\bibinfo {year} {2015})}\BibitemShut {NoStop}%
\bibitem [{\citenamefont {Hofheinz}\ \emph {et~al.}(2006)\citenamefont
  {Hofheinz}, \citenamefont {Jehl}, \citenamefont {Sanquer}, \citenamefont
  {Molas}, \citenamefont {Vinet},\ and\ \citenamefont
  {Deleonibus}}]{hofheinz2006simple}%
  \BibitemOpen
  \bibfield  {author} {\bibinfo {author} {\bibfnamefont {M.}~\bibnamefont
  {Hofheinz}}, \bibinfo {author} {\bibfnamefont {X.}~\bibnamefont {Jehl}},
  \bibinfo {author} {\bibfnamefont {M.}~\bibnamefont {Sanquer}}, \bibinfo
  {author} {\bibfnamefont {G.}~\bibnamefont {Molas}}, \bibinfo {author}
  {\bibfnamefont {M.}~\bibnamefont {Vinet}}, \ and\ \bibinfo {author}
  {\bibfnamefont {S.}~\bibnamefont {Deleonibus}},\ }\href@noop {} {\bibfield
  {journal} {\bibinfo  {journal} {Appl. Phys. Lett.}\ }\textbf {\bibinfo
  {volume} {89}},\ \bibinfo {pages} {143504} (\bibinfo {year}
  {2006})}\BibitemShut {NoStop}%
\bibitem [{\citenamefont {West}\ \emph {et~al.}()\citenamefont {West},
  \citenamefont {Hensen}, \citenamefont {Jouan}, \citenamefont {Tanttu},
  \citenamefont {Yang}, \citenamefont {Rossi}, \citenamefont {Gonzalez-Zalba},
  \citenamefont {Hudson}, \citenamefont {Morello}, \citenamefont {Reilly},\
  and\ \citenamefont {Dzurak}}]{DzurakReflecto}%
  \BibitemOpen
  \bibfield  {author} {\bibinfo {author} {\bibfnamefont {A.}~\bibnamefont
  {West}}, \bibinfo {author} {\bibfnamefont {B.}~\bibnamefont {Hensen}},
  \bibinfo {author} {\bibfnamefont {A.}~\bibnamefont {Jouan}}, \bibinfo
  {author} {\bibfnamefont {T.}~\bibnamefont {Tanttu}}, \bibinfo {author}
  {\bibfnamefont {C.}~\bibnamefont {Yang}}, \bibinfo {author} {\bibfnamefont
  {A.}~\bibnamefont {Rossi}}, \bibinfo {author} {\bibfnamefont
  {M.}~\bibnamefont {Gonzalez-Zalba}}, \bibinfo {author} {\bibfnamefont
  {F.}~\bibnamefont {Hudson}}, \bibinfo {author} {\bibfnamefont
  {A.}~\bibnamefont {Morello}}, \bibinfo {author} {\bibfnamefont
  {D.}~\bibnamefont {Reilly}}, \ and\ \bibinfo {author} {\bibfnamefont
  {A.}~\bibnamefont {Dzurak}},\ }\href@noop {} {\bibinfo  {journal}
  {arXiv:1809.01864}\ }\BibitemShut {NoStop}%
\bibitem [{\citenamefont {Pakkiam}\ \emph {et~al.}()\citenamefont {Pakkiam},
  \citenamefont {Timofeev}, \citenamefont {House}, \citenamefont {Kobayashi},
  \citenamefont {Koch}, \citenamefont {Rogge},\ and\ \citenamefont
  {Simmons}}]{SimmonsReflecto}%
  \BibitemOpen
\bibfield  {journal} {  }\bibfield  {author} {\bibinfo {author} {\bibfnamefont
  {P.}~\bibnamefont {Pakkiam}}, \bibinfo {author} {\bibfnamefont {A.~V.}\
  \bibnamefont {Timofeev}}, \bibinfo {author} {\bibfnamefont {M.}~\bibnamefont
  {House}, \bibfnamefont {M.G.~Hogg}}, \bibinfo {author} {\bibfnamefont
  {T.}~\bibnamefont {Kobayashi}}, \bibinfo {author} {\bibfnamefont
  {M.}~\bibnamefont {Koch}}, \bibinfo {author} {\bibfnamefont {S.}~\bibnamefont
  {Rogge}}, \ and\ \bibinfo {author} {\bibfnamefont {M.}~\bibnamefont
  {Simmons}},\ }\href@noop {} {\bibinfo  {journal} {arXiv:1809.01802}\
  }\BibitemShut {NoStop}%
\bibitem [{\citenamefont {Nakajima}\ \emph {et~al.}(2017)\citenamefont
  {Nakajima}, \citenamefont {Delbecq}, \citenamefont {Otsuka}, \citenamefont
  {Stano}, \citenamefont {Amaha}, \citenamefont {Yoneda}, \citenamefont
  {Noiri}, \citenamefont {Kawasaki}, \citenamefont {Takeda}, \citenamefont
  {Allison} \emph {et~al.}}]{nakajima2017robust}%
  \BibitemOpen
\bibfield  {journal} {  }\bibfield  {author} {\bibinfo {author} {\bibfnamefont
  {T.}~\bibnamefont {Nakajima}}, \bibinfo {author} {\bibfnamefont {M.~R.}\
  \bibnamefont {Delbecq}}, \bibinfo {author} {\bibfnamefont {T.}~\bibnamefont
  {Otsuka}}, \bibinfo {author} {\bibfnamefont {P.}~\bibnamefont {Stano}},
  \bibinfo {author} {\bibfnamefont {S.}~\bibnamefont {Amaha}}, \bibinfo
  {author} {\bibfnamefont {J.}~\bibnamefont {Yoneda}}, \bibinfo {author}
  {\bibfnamefont {A.}~\bibnamefont {Noiri}}, \bibinfo {author} {\bibfnamefont
  {K.}~\bibnamefont {Kawasaki}}, \bibinfo {author} {\bibfnamefont
  {K.}~\bibnamefont {Takeda}}, \bibinfo {author} {\bibfnamefont
  {G.}~\bibnamefont {Allison}},  \emph {et~al.},\ }\href@noop {} {\bibfield
  {journal} {\bibinfo  {journal} {Phys. Rev. Lett.}\ }\textbf {\bibinfo
  {volume} {119}},\ \bibinfo {pages} {017701} (\bibinfo {year}
  {2017})}\BibitemShut {NoStop}%
\bibitem [{\citenamefont {Yang}\ \emph {et~al.}(2013)\citenamefont {Yang},
  \citenamefont {Rossi}, \citenamefont {Ruskov}, \citenamefont {Lai},
  \citenamefont {Mohiyaddin}, \citenamefont {Lee}, \citenamefont {Tahan},
  \citenamefont {Klimeck}, \citenamefont {Morello},\ and\ \citenamefont
  {Dzurak}}]{yang2013spin}%
  \BibitemOpen
  \bibfield  {author} {\bibinfo {author} {\bibfnamefont {C.}~\bibnamefont
  {Yang}}, \bibinfo {author} {\bibfnamefont {A.}~\bibnamefont {Rossi}},
  \bibinfo {author} {\bibfnamefont {R.}~\bibnamefont {Ruskov}}, \bibinfo
  {author} {\bibfnamefont {N.}~\bibnamefont {Lai}}, \bibinfo {author}
  {\bibfnamefont {F.}~\bibnamefont {Mohiyaddin}}, \bibinfo {author}
  {\bibfnamefont {S.}~\bibnamefont {Lee}}, \bibinfo {author} {\bibfnamefont
  {C.}~\bibnamefont {Tahan}}, \bibinfo {author} {\bibfnamefont
  {G.}~\bibnamefont {Klimeck}}, \bibinfo {author} {\bibfnamefont
  {A.}~\bibnamefont {Morello}}, \ and\ \bibinfo {author} {\bibfnamefont
  {A.}~\bibnamefont {Dzurak}},\ }\href@noop {} {\bibfield  {journal} {\bibinfo
  {journal} {Nat. Commun.}\ }\textbf {\bibinfo {volume} {4}},\ \bibinfo {pages}
  {2069} (\bibinfo {year} {2013})}\BibitemShut {NoStop}%
\bibitem [{\citenamefont {Macklin}\ \emph {et~al.}(2015)\citenamefont
  {Macklin}, \citenamefont {O’Brien}, \citenamefont {Hover}, \citenamefont
  {Schwartz}, \citenamefont {Bolkhovsky}, \citenamefont {Zhang}, \citenamefont
  {Oliver},\ and\ \citenamefont {Siddiqi}}]{macklin2015near}%
  \BibitemOpen
  \bibfield  {author} {\bibinfo {author} {\bibfnamefont {C.}~\bibnamefont
  {Macklin}}, \bibinfo {author} {\bibfnamefont {K.}~\bibnamefont {O’Brien}},
  \bibinfo {author} {\bibfnamefont {D.}~\bibnamefont {Hover}}, \bibinfo
  {author} {\bibfnamefont {M.}~\bibnamefont {Schwartz}}, \bibinfo {author}
  {\bibfnamefont {V.}~\bibnamefont {Bolkhovsky}}, \bibinfo {author}
  {\bibfnamefont {X.}~\bibnamefont {Zhang}}, \bibinfo {author} {\bibfnamefont
  {W.}~\bibnamefont {Oliver}}, \ and\ \bibinfo {author} {\bibfnamefont
  {I.}~\bibnamefont {Siddiqi}},\ }\href@noop {} {\bibfield  {journal} {\bibinfo
   {journal} {Science}\ }\textbf {\bibinfo {volume} {350}},\ \bibinfo {pages}
  {307} (\bibinfo {year} {2015})}\BibitemShut {NoStop}%
\bibitem [{\citenamefont {Petit}\ \emph {et~al.}(2018)\citenamefont {Petit},
  \citenamefont {Boter}, \citenamefont {Eenink}, \citenamefont {Droulers},
  \citenamefont {Tagliaferri}, \citenamefont {Li}, \citenamefont {Franke},
  \citenamefont {Singh}, \citenamefont {Clarke}, \citenamefont {Schouten},
  \citenamefont {Dobrovitski}, \citenamefont {Vandersypen},\ and\ \citenamefont
  {Veldhorst}}]{PhysRevLett.121.076801}%
  \BibitemOpen
  \bibfield  {author} {\bibinfo {author} {\bibfnamefont {L.}~\bibnamefont
  {Petit}}, \bibinfo {author} {\bibfnamefont {J.~M.}\ \bibnamefont {Boter}},
  \bibinfo {author} {\bibfnamefont {H.~G.~J.}\ \bibnamefont {Eenink}}, \bibinfo
  {author} {\bibfnamefont {G.}~\bibnamefont {Droulers}}, \bibinfo {author}
  {\bibfnamefont {M.~L.~V.}\ \bibnamefont {Tagliaferri}}, \bibinfo {author}
  {\bibfnamefont {R.}~\bibnamefont {Li}}, \bibinfo {author} {\bibfnamefont
  {D.~P.}\ \bibnamefont {Franke}}, \bibinfo {author} {\bibfnamefont {K.~J.}\
  \bibnamefont {Singh}}, \bibinfo {author} {\bibfnamefont {J.~S.}\ \bibnamefont
  {Clarke}}, \bibinfo {author} {\bibfnamefont {R.~N.}\ \bibnamefont
  {Schouten}}, \bibinfo {author} {\bibfnamefont {V.~V.}\ \bibnamefont
  {Dobrovitski}}, \bibinfo {author} {\bibfnamefont {L.~M.~K.}\ \bibnamefont
  {Vandersypen}}, \ and\ \bibinfo {author} {\bibfnamefont {M.}~\bibnamefont
  {Veldhorst}},\ }\href@noop {} {\bibfield  {journal} {\bibinfo  {journal}
  {Phys. Rev. Lett.}\ }\textbf {\bibinfo {volume} {121}},\ \bibinfo {pages}
  {076801} (\bibinfo {year} {2018})}\BibitemShut {NoStop}%
\end{thebibliography}%

\end{document}